\documentclass[twocolumn,showpacs,amssymb]{revtex4}
\usepackage{graphicx}
\usepackage{dcolumn}
\usepackage{bm}
\usepackage{epsfig}
\usepackage{amsmath}

\begin{document}

\title{Testing Einstein's Weak Equivalence Principle With Gravitational Waves}

\author{Xue-Feng Wu$^{1,2\ast}$, He Gao$^{3\ast}$,
Jun-Jie Wei$^{1}$, Peter M{\'e}sz{\'a}ros$^{4,5,6}$, Bing Zhang$^{7}$, Zi-Gao Dai$^{8}$, Shuang-Nan Zhang$^{9,10}$ and Zong-Hong Zhu$^{3}$}

\affiliation{$^1$ Purple Mountain Observatory, Chinese Academy of Sciences, Nanjing 210008, China \\
$^2$ Joint Center for Particle, Nuclear Physics and Cosmology, Nanjing
University-Purple Mountain Observatory, Nanjing 210008, China\\
$^3$ Department of Astronomy, Beijing Normal University, Beijing 100875, China\\
$^4$ Department of Astronomy and Astrophysics, Pennsylvania State University, 525 Davey Laboratory, University Park, PA 16802, USA\\
$^5$ Department of Physics, Pennsylvania State University, 104 Davey Laboratory, University Park, PA 16802, USA\\
$^6$ Center for Particle and Gravitational Astrophysics, Institute for Gravitation and the Cosmos, Pennsylvania State University, 525 Davey
Laboratory, University Park, PA 16802, USA\\
$^7$ Department of Physics and Astronomy, University of Nevada Las Vegas, Las Vegas, NV 89154, USA\\
$^8$ School of Astronomy and Space Science, Nanjing University, Nanjing 210093, China\\
$^{9}$ Laboratory for Particle Astrophysics, Institute of High Energy Physics, Beijing 100049, China\\
$^{10}$ National Astronomical Observatories, Chinese Academy Of Sciences, Beijing 100012, China.\\
$^\ast$Electronic address: xfwu@pmo.ac.cn;gaohe@bnu.edu.cn}

\date{\today}


\begin{abstract}
A conservative constraint on the Einstein Weak Equivalence Principle (WEP) can be obtained under the
assumption that the observed time delay between correlated particles from astronomical sources is
dominated by the gravitational fields through which they move. Current limits on the WEP are mainly
based on the observed time delays of photons with different energies. It is highly desirable
to develop more accurate tests that include the gravitational wave (GW) sector. The detection by
the advanced LIGO/VIRGO systems of gravitational waves will provide attractive candidates for
constraining the WEP, extending the tests to gravitational interactions, with
potentially higher accuracy. Considering the capabilities of the advanced LIGO/VIRGO network and
the source direction uncertainty, we show that the joint detection of GWs and electromagnetic
signals could probe the WEP to an accuracy down to $10^{-10}$, which is one order of
magnitude tighter than previous limits, and seven orders
of magnitude tighter than the multi-messenger (photons and neutrinos) results by supernova 1987A.
\end{abstract}

\maketitle

\section{Introduction}

Albert Einstein's Weak Equivalence Principle (WEP) is one of the main cornerstones of general relativity
as well as of many other gravitational theories. One statement of the WEP is that any
freely falling, uncharged test body will follow a trajectory independent of its internal composition
and structure. It implies that any two different species of massless
(or negligible rest mass) neutral particles, or two particles of same species with different energies,
if emitted simultaneously from the same source and traveling through the same gravitational fields,
should reach us at the same time \cite{will06,will14}.
By measuring how closely in time the two different particles arrive, one can test the accuracy of the WEP
through the Shapiro (gravitational) time delay effect \cite{shapiro64}. In practice, all metric
theories of gravity incorporating the WEP predict that all test particles must follow identical trajectories and
undergo the same Shapiro time delay. In other words, as long as the WEP is valid, all metric
theories predict $\gamma_{1}=\gamma_{2}\equiv\gamma$, where $\gamma$ is the parametrized post-Newtonian (PPN) parameter
($\gamma$ denotes how much space curvature is provided by unit rest mass of the objects along or near the path of the particles \cite{will06,will14})
and the subscripts represent two different particles. In this case, the WEP validity
can be characterized by limits on the differences of $\gamma$ value for different test particles (see, e.g.,
Refs.~\cite{krauss88,longo88,sivaram99,gao15,wei15,Tingay16,wei16}).

Any possible violation of the WEP would have far-reaching consequences for mankind's view of nature,
so it is important to extend the tests of its validity by making use of the panoply of new types
of astronomical signals being brought to the fore in the multi-messenger era.
So far, tests of the WEP through the relative differential variations of the $\gamma$ values
have been made using the emissions from supernova 1987A \cite{krauss88,longo88}, gamma-ray bursts
(GRBs) \cite{sivaram99,gao15}, fast radio bursts (FRBs) \cite{wei15,Tingay16}, and TeV blazars
\cite{wei16}. Particularly, assuming that the observed time delay between different
frequency photons from FRBs are caused mainly by the gravitational potential of the Milky Way,
Ref.~\cite{wei15} set the most stringent limits to date on $\gamma$ differences, yielding $\sim 10^{-8}$.
Even more encouragingly, the most recent studies \cite{nusser16+zhang16} show that the constraints
on the WEP accuracy from FRBs can be further improved by a few orders of magnitude when taking into
account the gravitational potential fluctuations of the large scale structure, rather than the
Milky Way's gravity.
In addition, the discovery of a triple system \cite{Ransom14}, made of a
millisecond pulsar PSR J0337+1715 and two white dwarves, has recently provided a new interesting
test of the equivalence principle. The very large difference in the gravitational
binding energies of the pulsar and the white dwarf makes this system very promising
on the equivalence principle test.

Although the tests on the WEP have reached high precision, most of
the tests rely on the relative arrival time delays of (exclusively) photons
with different energies.
The first and only WEP test with different species of particles was the measurement of the time delay
between the photons and neutrinos from supernova 1987A \cite{krauss88,longo88}. It was shown
that the $\gamma$ values of photons and neutrinos are equal to an accuracy of approximately 0.34\%.
New multi-messenger signals exploiting different emission channels are essential for testing the WEP
to a higher accuracy.

Recently, the Laser Interferometer Gravitational-wave Observatory
(LIGO) team report their discovery of the first gravitational wave (GW) source, GW 150914
\cite{abbott16a}, opening a brand new channel for studying the Universe, which could lead to
breakthroughs in both fundamental physics and astrophysics. In fact, the next generation of
gravitational detectors, including the advanced LIGO, advanced VIRGO and KAGRA, appear poised to
detect a plethora of increasingly sophisticated gravitational wave (GW) signals in the very near
future \cite{abbott09,acernese08,kuroda10,acernese15,aasi15}. Phenomenologically, one may
treat the GWs with different frequencies as different gravitons to test the WEP. In extending the
constraints on $\Delta \gamma$, tests of the WEP using GW measurements become more robust against
various assumptions, since, e.g. GWs do not suffer absorption or scattering along their path,
in contrast to photons. In the following, we illustrate the progress that can be expected in testing
the WEP with the reported/future GW observations.

\section{Description of the Method}

The Shapiro time delay effect \cite{shapiro64} causes the time interval for particles to pass
through a given distance to be longer in the presence of a gravitational potential $U(r)$ by
\begin{equation}
\Delta t_{\rm gra}=-\frac{1+\gamma}{c^3}\int_{r_e}^{r_o}~U(r)dr\;,
\end{equation}
where $\gamma$ is a PPN parameter, $r_{o}$ and $r_{e}$ correspond to locations of observation
and the source of particle emission.

Assuming that the observed time delays $(\Delta t_{\rm obs})$ between correlated particles
from the same astronomical source are mainly caused by the gravitational potential of the Milky Way,
and adopting the Keplerian potential for the Milky Way, we have \cite{longo88,misner73}
\begin{equation}
\begin{split}
\Delta t_{\rm obs}>\Delta t_{\rm gra}=\Delta \gamma \frac{GM_{\rm MW}}{c^{3}} \times\qquad\qquad\qquad\qquad\qquad\\
\ln \left\{ \frac{ \left[d+\left(d^{2}-b^{2}\right)^{1/2}\right] \left[r_{G}+s_{\rm n}\left(r_{G}^{2}-b^{2}\right)^{1/2}\right] }{b^{2}} \right\}\;,
\end{split}
\label{eq:gammadiff}
\end{equation}
where $\Delta \gamma$ is the difference between the $\gamma$ values for different test particles,
$M_{\rm MW}\simeq6\times10^{11}M_{\odot}$ is the Milky Way mass \cite{mcmillan11,footnote},
$d$ represents the distance from the source to the center of the Milky Way (if the source is of
extra-galactic or cosmological origin, $d$ is approximated as the distance from the source to
Earth), $r_{G}\simeq8$ kpc is the distance from the Sun to the center of the Milky Way, $b$ denotes
the impact parameter of the particle paths relative to the Milky Way center, and $s_{\rm n}=\pm1$
is the sign of the correction of the source direction.
If the source is located along the direction of the Galactic center, $s_{\rm n}=+1$. While, $s_{\rm n}=-1$
corresponds to the source located along the direction pointing away from the Galactic center. Note that the impact
parameter $b$ is on the order of the distance of the Sun from the Galactic center, i.e.,
$b\leq r_{G}$. With Equation \ref{eq:gammadiff}, one can constrain the WEP by putting a strict limit
on the differences of $\gamma$ value \cite{krauss88,longo88,sivaram99,gao15,wei15,Tingay16,wei16}.

We notice that although the method adopted in this work can provide severe constraints on the accuracy of the WEP, which is one of the important postulates of GR, it can not be directly used to distinguish
between specific gravity theories, such as GR and its alternatives.  Many precise methods have been
devised to test the accuracy of GR through the measurement of the absolute value of $\gamma$ based
on the fact that GR predicts $\gamma=1$ (see Ref. \cite{will14} for a recent review). However, it is worth pointing out that $\gamma=1$ is not a
sufficient condition to identify general relativity, since it is not the only theory that predicts
$\gamma=1$ \cite{will14}. Thus, further investigations would be essential for
developing more accurate tests of the WEP and for distinguishing between GR and other alternative
gravity theories.

\begin{figure}[h]
\epsfig{figure=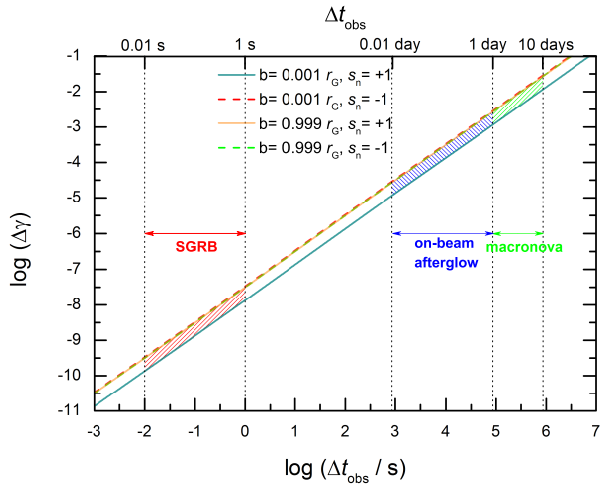,width=3.5truein}
\vskip-0.1in
\caption{Expected limits on the differences of the $\gamma$ values between
the GW signals and the photons for various types of EM counterparts.
The vertical lines correspond to different characteristic times.}
\end{figure}

\section{WEP test using GW signals}

The process of compact binary coalescence (CBC; either neutron star (NS) binary, black hole
(BH) binary or NS-BH binary) provides the primary targets for the second generation of GW detectors,
such as the advanced LIGO/VIRGO \cite{abbott09,acernese08,kuroda10,acernese15,aasi15}. The first
reported GW detection, GW 150914, is a BH-BH merger with two BH masses $36^{+5}_{-4} \rm{M_{\odot}}$
and $29^{+4}_{-4} \rm{M_{\odot}}$, respectively \cite{abbott16a}. Of significant interest for
CBC GW detections is the fact that some
relevant fundamental physics postulates, including the WEP, may be constrained using gravitational radiation alone
\cite{will14,will98}. This could be done exploiting the fact that the frequency of the
gravitational radiation sweeps from low frequencies at the initial
moment of observation (in-spiral phase) to a higher frequency at the
final moment (coalescence phase). Note that the GW frequency eventually
saturates to a constant value in the vicinity of the light-ring.
The amplitude, however, decreases monotonically after
reaching its peak at the light-ring.
Any WEP violation will cause a distortion of the observed phasing of
the waves, and would result in a shorter (or longer) than expected
overall time of passage of a given number of cycles.
It is worth pointing out that there are many effects that can change
the lifespan of a waveform: spin corrections, eccentricity, spin
precession, etc. However, it is difficult to disentangle these effects from a WEP
violation. Our upper limits on the WEP accuracy are based
on very conservative estimates of the observed time delay
(i.e., the whole time delay is assumed to be caused by the WEP violation).
In fact, the inclusion of contributions from the neglected effects
in the observed waveform could improve the limits on WEP to some degree.
In this case, since no EM counterparts are required, all CBC GW detections would be relevant.

For instance, the signal of GW 150914 increases in frequency and amplitude in about 8 cycles
(over 0.2 s) from 35 to 150 Hz, where the amplitude reaches a maximum \cite{abbott16a}. Considering
the localization information of GW 150914, we could tighten the limit on the WEP to
$\Delta \gamma \sim 10^{-9}$.

More recently, the Fermi Gamma-Ray Burst Monitor (GBM) team reported that GBM observations at the time of GW150914 reveal the presence of a
weak transient source above 50 keV, 0.4 s after the GW event was detected, with a false alarm probability of 0.0022
\cite{connaughton16}. If this is indeed the EM counterpart of GW150914 (see possible interpretations in \cite{zhangb16+loeb16}), with the aforementioned method, we could
further extend the WEP test with GWs and photons, setting a severe limit on WEP to an accuracy of $10^{-8}$, five
orders of magnitude tighter than the results set by the photons and neutrinos from supernova 1987A.

Besides BH-BH mergers, GW signals from binary NSs and NS-BH mergers are also expected to be detected in the near future \cite{abadie10}, for which a variety of detectable electromagnetic (EM) counterparts have been widely discussed
\cite{eichler89,li98,metzger12}, including
the following representative cases: the prompt short GRB emission, the afterglow emission of the
on-beam ultra-relativistic outflows, and the macronova/kilonova emission of the sub-relativistic
r-process material ejected during the merger. For NS-NS mergers, if the merger product is a massive millisecond pulsar
instead of a BH, the detectable EM signatures from the system
become much richer and brighter (see Ref. \cite{zhang13+gao13} for details). Joint detections of GW/EM signals,
once achieved, could be used to give important constraints on the WEP.

Consider the case of a joint detection of GW/EM signals from a NS-NS or NS-BH coalescence event in
the advanced LIGO/VIRGO era. Since the sky and binary orientation averaged sensitivity of the advanced
LIGO/VIRGO network for CBC is of the order of $\sim100$ Mpc \cite{abbott09,acernese08,kuroda10,acernese15,aasi15}, here we assume the
distance from the GW source to the Earth to be $d=200$ Mpc. It is worth pointing out that the constraints
on the WEP are not greatly affected by the source distance uncertainty (see Ref.~\cite{wei15} for
more explanations). To account for the source direction uncertainty, and based on the fact that the
impact parameter $b\leq r_{G}$, here we present four extreme cases by assuming $b=0.001r_{G}$ and
$s_{\rm n}=+1$, $b=0.001r_{G}$ and $s_{\rm n}=-1$, $b=0.999r_{G}$ and $s_{\rm n}=+1$, and $b=0.999r_{G}$
and $s_{\rm n}=-1$, respectively. The real results should lie within the range circumscribed by these
extreme cases.

Regarding the EM counterpart of the GW detection, suppose we are lucky to detect all the promising
emission types, e.g. the short GRB prompt emission, the on-beam GRB afterglow emission and the
macronova emission. Recently, Ref.~\cite{li16} discussed the time lags between the GW signal and
all these EM counterparts in some detail, and suggested that the time delay $\Delta t_{\rm obs}$ is
expected to be of the order of $\sim$ 0.01--1 s (short GRB), 0.01--1 day (on-beam afterglow), or
1--10 days (macronova), respectively. With these expected time delays and with the location information
in hand, we would be able to set bounds on the WEP from Equation~(\ref{eq:gammadiff}). The expected
constraints on the differences of the $\gamma$ values are shown in Figure~1. It has been suggested
that the macronova emission may be the most frequently-detectable EM signal of the coalescence events
\cite{li98,metzger12}. If the macronova emission is detected at $\Delta t_{\rm obs}\sim1$ day after the merger,
a strict limit on the WEP will be $\Delta \gamma <10^{-3}$. One can see from this plot that much more
severe constraints would be achieved ($\sim$ $10^{-3}$--$10^{-5}$ or $10^{-8}$--$10^{-10}$) if the
EM counterpart is an on-beam afterglow or a short GRB. Note that the compact binary coalescence and
the EM counterpart do not occur at the same time, since $\Delta t_{\rm obs}$ has a contribution from
the intrinsic emission time lag ($\Delta t_{\rm lag}$) between the photons and the GW signals. Here
we take $\Delta t_{\rm lag}=0$ to give a conservative estimate of the WEP. More severe constraints
could be achieved with a better understanding of the nature of  $\Delta t_{\rm lag}$ allowing one
to remove its contribution from $\Delta t_{\rm obs}$. On the other hand, it should be underlined
that these upper limits are based on very conservative estimates of the gravitational potential of the
Milky Way. If the gravitational potential fluctuations from the intervening large scale structures are
taken into considered, our constraint results would be further improved by orders of magnitude
\cite{nusser16+zhang16}.

\section{Summary and discussion}

In conclusion, we show that new WEP tests can be carried out with potentially much higher
accuracy in the GW era.
For all kinds of CBC GW detections, regardless of whether EM counterparts are detected or not, we can always
use GWs with different frequencies to give stringent constraints on the accuracy of the WEP. Taking GW 150914 as
an example, it takes less than one second for the GW signals emitted from lower frequency to higher frequency
where the signal amplitude reaches a maximum (e.g., 35 Hz to 150 Hz), resulting in a tightening of the
limit on the WEP to approximately $10^{-9}$, which is as good as the current
most stringent results from FRBs \cite{wei15,Tingay16}.

Once EM counterparts of the GW signals are firmly detected, an interesting WEP test could be performed
by using the measured time delay between the GWs and any associated photons. Also taking GW 150914
as an example, if the claimed short GRB, GW150914-GBM, is indeed the EM counterpart of GW150914, a severe
limit on WEP could be set to an accuracy of $10^{-8}$, five orders of magnitude tighter than the results set
by the photons and neutrinos from supernova 1987A \cite{krauss88,longo88}.

Finally, considering the capabilities of the advanced LIGO/VIRGO network and the source direction
uncertainty, we found that for the expected GW detection from NS-NS/BH mergers, if the prompt short GRB
emission and/or its afterglow emission is detected, a stringent limit on the WEP could be set at the
level of $\Delta \gamma < (10^{-8}$--$10^{-10})$ (prompt) or $\sim$ $10^{-3}$--$10^{-5}$ (afterglow).
Due to the low detection rates of GRB- accompanied GW signals, the first positively identified
electromagnetic counterpart of a GW signal is very likely to be a macronova.  If the macronova emission
is detected at $\Delta t_{\rm obs}\sim1$ day after the merger, a strict limit on the WEP will be
$\Delta \gamma <10^{-3}$.

In sum, the main result of this paper is to propose a method to test WEP,
which can be applied when future robust GW/EM associations become available.
For GW 150914, we have applied our method to the available data
(the GW data and the putative EM signal following the GW signal) and derived some
stringent limit on WEP, not achievable by previous analyses.
There are astrophysical uncertainties in applying our method. Examples of such astrophysical uncertainties are, e.g. the
distance of a purely GW-detected sources such as GW 150914; the astrophysical time lags between
EM and GW emission mentioned in Ref. \cite{li16}; the detection of more than one EM emission component
of a short GRB; etc. Such astrophysical uncertainties, however, will certainly diminish in time,
with improving EM instruments and observations, the addition of further GW detectors at
different Earth locations, etc.

\vskip 0.1in
\noindent{\bf Acknowledgements:}
We are grateful to the anonymous referees for insightful comments.
We also thank Dr. Xi-Long Fan, who can not be a co-author
due to some restrictions as a member of the LIGO collaboration, for extensive
discussions and actual contribution to this manuscript.
This work is partially supported by the National Basic Research Program (``973'' Program)
of China (Grants 2014CB845800 and 2013CB834900), the National Natural Science
Foundation of China (grants Nos. 11322328, 11433009, 11543005, 11573014, and 11303009),
the Youth Innovation Promotion Association (2011231), the Strategic Priority Research Program
``The Emergence of Cosmological Structures'' (Grant No. XDB09000000) of the Chinese Academy of Sciences, the Natural Science
 Foundation of Jiangsu Province (Grant No. BK20161096), and NASA NNX 13AH50G, 14AF85G and 15AK85G.


\begin{thebibliography}{99}
\bibitem{footnote} One can see from Equation~(\ref{eq:gammadiff}) that
the limit of $\Delta \gamma$ is inversely proportional to the value of the
Milky Way mass. Currently, the mass of the Milky Way $M_{\rm MW}$ is well constrained
and the uncertainty of $M_{\rm MW}$ is just about 10\% \cite{mcmillan11}.
Hence, the limit of $\Delta \gamma$ would not be greatly affected by the uncertainty of $M_{\rm MW}$.
\end{thebibliography}
\end{document}